\documentclass[12pt]{article}
\usepackage{axodraw}
\usepackage{epsfig}
\def\e{{\rm e}}
\newcommand{\be}{\begin{equation}}
\newcommand{\ee}{\end{equation}}
\newcommand{\bea}{\begin{eqnarray}}
\newcommand{\eea}{\end{eqnarray}}

\newcommand{\gm}{\gamma}
\newcommand{\Gm}{\Gamma}

\newcommand{\ep}{\epsilon}

\newcommand{\dd}{\mbox{d}}

\newcommand{\nn}{\nonumber}

\newcommand{\Li}[2]{{\mbox{Li}}_{#1}\left(#2\right)}

\begin{document}
\parindent=1.5pc

\begin{titlepage}
\begin{flushright}
DESY 04--94\\
June 2004
\end{flushright}
\vspace{1.cm}
\bigskip
\begin{center}
{{\bf
Analytical Evaluation of Dimensionally Regularized Massive
On-Shell Double Boxes
} \\
\vglue 5pt
\vglue 1.0cm
{\large G.~Heinrich}\\
\baselineskip=14pt
\vspace{2mm}
{\normalsize II. Institut f\"ur Theoretische Physik, Universit\"at Hamburg,}\\
{\normalsize Luruper Chaussee 149, 22761 Hamburg, Germany}\\
\baselineskip=14pt
\vspace{2mm}
and\\
\baselineskip=14pt
\vspace{2mm}
{ {\large V.A. Smirnov} }\\
\baselineskip=14pt
\vspace{2mm}
{Nuclear Physics Institute of Moscow State University,\\
Moscow 119992, Russia}
\baselineskip=14pt
\vspace{2mm}
\vglue 0.8cm
{\bf Abstract}}
\end{center}
\vglue 0.3cm
{\rightskip=3pc
 \leftskip=3pc
\noindent
The method of Mellin--Barnes representation is used to calculate
dimensionally regularized massive on-shell double box {}Feynman diagrams
contributing to Bhabha scattering at two loops.
%\ldots
\vglue 0.8cm}
\end{titlepage}

\section{Introduction}

The method of Mellin--Barnes representation has turned
out to be very successful for the analytical evaluation of
two-loop Feynman diagrams with four external lines within
dimensional regularization \cite{dimreg}.
The problem of the evaluation of massless on-shell double boxes
was solved in \cite{MB,GTGR0ATT,SV,AGORT}, with
multiple subsequent applications to the calculation of two-loop
scattering amplitudes in gauge theories\,\cite{appl}.
First calculations of massless on-shell four-point three-loop Feynman
integrals were done in \cite{3b}.

Complete algorithms for the evaluation of massless on-shell
double boxes with one leg off shell were also constructed.
Master integrals were calculated using the MB representation
(first results in \cite{S2})
and the method of differential equations (DE) \cite{DE},
where a systematic evaluation was described in \cite{LGR2}.
The reduction to master integrals was done using a Laporta's idea
\cite{LGR1} in \cite{LGR2}. All results are expressed in terms of
two-dimensional harmonic polylogarithms  \cite{LGR2} which
generalize harmonic polylogarithms (HPL) \cite{HPL}.
Let us stress that this very combination of reduction based on \cite{LGR1,LGR2}
and DE was successfully applied in numerous calculations, e.g. to
various classes of vertex diagrams \cite{appl-vert1,appl-vert2}.

Another important class of Feynman integrals with one more
parameter, with respect to the massless on--shell double boxes,
are massive on-shell double boxes relevant to Bhabha scattering at NNLO.
\begin {figure}[htbp]
\begin{picture}(400,100)(-15,-20)

\Line(-15,0)(0,0)
\Line(-15,50)(0,50)
\Line(115,0)(100,0)
\Line(115,50)(100,50)
\Line(0,0)(50,0)
\Line(50,0)(100,0)
\Line(100,50)(50,50)
\Line(50,50)(0,50)

\DashLine(0,50)(0,0){3}
\DashLine(50,0)(50,50){3}
\DashLine(100,0)(100,50){3}
%%%%%%%%%%%%%
\Vertex(0,0){1.5}
\Vertex(50,0){1.5}
\Vertex(100,0){1.5}
\Vertex(0,50){1.5}
\Vertex(50,50){1.5}
\Vertex(100,50){1.5}

\Text(50,-15)[]{(a)}
%%%%%%%%%%%%%%%%%%%%%%%%%%%%%%%%%%%%%%%%%%%%%%%%%%%%%%%%%

\Line(135,0)(150,0)
\Line(135,50)(150,50)
\Line(265,0)(250,0)
\Line(265,50)(250,50)

\Line(150,0)(200,0)
\DashLine(200,0)(250,0){3}
\Line(150,50)(200,50)
\DashLine(200,50)(250,50){3}

\DashLine(150,50)(150,0){3}
\Line(200,0)(200,50)
\Line(250,0)(250,50)

\Vertex(150,0){1.5}
\Vertex(200,0){1.5}
\Vertex(250,0){1.5}
\Vertex(150,50){1.5}
\Vertex(200,50){1.5}
\Vertex(250,50){1.5}

\Text(200,-15)[]{(b)}

%%%%%%%%%%%%%%%%%%%
\Line(285,0)(300,0)
\Line(285,50)(300,50)
\Line(415,0)(400,0)
\Line(415,50)(400,50)
\Line(300,0)(350,0)
\Line(350,0)(400,0)
\Line(400,50)(350,50)
\Line(350,50)(300,50)

\DashLine(300,50)(300,0){3}
\DashLine(350,0)(400,50){3}
\DashLine(400,0)(350,50){3}
%%%%%%%%%%%%%
\Vertex(300,0){1.5}
\Vertex(350,0){1.5}
\Vertex(400,0){1.5}
\Vertex(300,50){1.5}
\Vertex(350,50){1.5}
\Vertex(400,50){1.5}

\Text(350,-15)[]{(c)}

\end{picture}
\caption{Massive on-shell double boxes: (a)~planar double box of the first type,
(b)~planar double box of the second type and (c)~non-planar double box.
The solid lines denote massive, the dotted lines massless particles.}
\label{2boxes}
\end{figure}
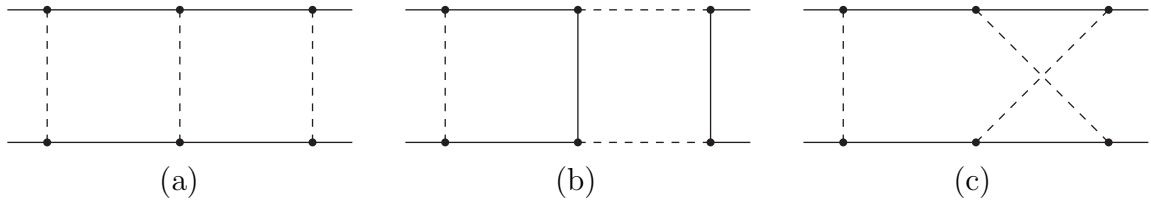
Bhabha scattering plays an important role as a luminosity monitor for
$e^+e^-$ colliders. Because of this phenomenological interest,
the two-loop corrections to Bhabha scattering have been studied
extensively in recent years.
In \cite{Bern:2000ie}, the complete two-loop virtual
QED corrections were calculated in the massless fermion approximation.
The infrared  singularity structure of this result was studied
in \cite{Glover:2001ev}.
The first result for a two-loop master double box
with nonzero electron mass was obtained in \cite{S3}.
Planar two-loop box diagrams  with one-loop insertion and
without neglecting the fermion masses were
calculated by DE in \cite{Bonciani:2003cj}. 
This work was continued to calculate
the subset of all two-loop diagrams involving a closed fermion loop
in \cite{Bonciani:2004gi}.
The  two-loop contribution which factorizes
into squares of one-loop diagrams was calculated in \cite{Fleischer:2002wa}.
Numerical results for the three master integrals
shown in Fig.\,\ref{2boxes} were given in \cite{baba} for Euclidean
points. Despite all these efforts, a complete two-loop differential
cross section is not available yet\,\cite{Gluza},
mainly because the most complicated
master integrals, the two-loop box diagrams, have not all been calculated
so far.

\section{Evaluating by Mellin-Barnes representation}

\subsection{Planar double boxes of the first type}

The general double box Feynman integral of the first type
(see Fig.~1a) takes the form
\bea
B_{PL,1}(a_1,\ldots,a_8;s,t,m^2;\ep) &=&
\int\int \frac{\dd^dk \, \dd^dl}{(k^2-m^2)^{a_1}[(k+p_1)^2)]^{a_2}
[(k+p_1+p_2)^2-m^2]^{a_3}}
\nn \\ && \hspace*{-50mm}
\times \frac{[(k+p_1+p_2+p_3)^2]^{-a_8}}{
[(l+p_1+p_2)^2-m^2)]^{a_4}[(l+p_1+p_2+p_3)^2]^{a_5}
(l^2-m^2)^{a_6} [(k-l)^2]^{a_7} }
\, ,
\label{2box}
\eea
where $s=(p_1+p_2)^2, \;  t=(p_2+p_3)^2$, and
$k$ and $l$ are the loop momenta corresponding to the left and the right box.
Usual prescriptions $k^2=k^2+i 0, \; s=s+i 0$, etc. are implied.
For convenience,
we consider the factor with $(k+p_1+p_2+p_3)^2$ corresponding
to the irreducible numerator
as an extra propagator but, really, we are interested only in the
non-positive integer values of $a_8$.

A first analytical result for the planar double box of the first
type, was obtained in \cite{S3} using
the following sixfold MB representation:
\bea
B_{PL,1}(a_1,\ldots,a_8;s,t,m^2;\ep)
 &=&
\frac{\left(i\pi^{d/2} \right)^2 (-1)^a}{
\prod_{j=2,4,5,6,7}\Gm(a_j) \Gm(4-a_{4567}-2\ep)(-s)^{a-4+2\ep}}
\nn \\ &&  \hspace*{-50mm}\times
\frac{1}{(2\pi i)^6} \int_{-i\infty}^{+i\infty}
\dd w \prod_{j=1}^5 \dd z_j
\left(\frac{m^2}{-s} \right)^{z_1+z_5}
\left(\frac{t}{s} \right)^{w}
\frac{\Gm(a_2 + w) \Gm(-w) \Gm(z_2 + z_4) \Gm(z_3 + z_4)}
{\Gm(a_1 + z_3 + z_4) \Gm(a_3 + z_2 + z_4)}
\nn \\ &&  \hspace*{-50mm}\times
\frac{  \Gm(4 - a_{13} - 2a_{28} - 2 \ep + z_2 + z_3)
  \Gm(a_{1238} - 2 + \ep + z_4 + z_5)\Gm(a_7 + w - z_4)
  }
{\Gm(4 - a_{46} - 2 a_{57} - 2 \ep - 2 w - 2 z_1 - z_2 - z_3)}
\nn \\ &&  \hspace*{-50mm}\times
\frac{\Gm(a_{4567} - 2 + \ep + w + z_1 - z_4)
  \Gm(a_8 - z_2 - z_3 - z_4) \Gm(-w - z_2 - z_3 - z_4)}
{\Gm(4 - a_{1238} - 2 \ep + w - z_4)\Gm(a_8 - w - z_2 - z_3 - z_4) }
\nn \\ &&  \hspace*{-50mm}\times
\frac{\Gm(a_5 + w + z_2 + z_3 + z_4)
  \Gm(2 - a_{567} - \ep - w - z_1 - z_2) }
  {\Gm(4 - a_{13} - 2 a_{28} - 2 \ep + z_2 + z_3 - 2 z_5)}
  \nn \\ &&  \hspace*{-50mm}\times
\Gm(2 - a_{457} - \ep - w - z_1 - z_3)
  \Gm(2 - a_{128} - \ep + z_2 - z_5) \Gm(2 - a_{238} - \ep + z_3 - z_5)
  \nn \\ &&  \hspace*{-50mm}\times
  \Gm(4 - a_{46} - 2 a_{57} - 2 \ep - 2 w - z_2 - z_3) \Gm(-z_1)\Gm(-z_5)
\, ,
\label{6MB}
\eea
where $a=a_{1\ldots 8}$, $a_{4567}=a_4+a_5+a_6+a_7, a_{13}=a_1+a_3$, etc.,
and integration contours are chosen in the standard way.

To evaluate  the master double box
$B_{PL,1}(1,\ldots,1,0;s,t,m^2;\ep)$
the standard procedure of taking residues and shifting contours \cite{MB}
was used \cite{S3}, with the goal to obtain a sum of integrals where
one may expand the integrands in a Laurent series in $\ep=(4-d)/2$
(where $d$ is the space-time dimension within dimensional
regularization \cite{dimreg}). Then one
can use the first and the second Barnes lemmas and their
corollaries to perform some of the MB integrations explicitly.
In the last integrations which usually carry a dependence
on the external variables, one closes the contour in the complex plane
and sums up the corresponding series.
(See \cite{S4} for details of the method.)

In \cite{Gluza} it was announced that
the massive on-shell double boxes with two reduced lines
have been calculated by DE, with a solution of reduction problem
by the method of \cite{LGR1,LGR2}.
However an extension of these results to all master integrals
meets problems. Differential equations of  third order
and higher are encountered there.

On the other hand, the method of MB representation can
certainly be applied to arbitrary massive on-shell double boxes.
To illustrate this point, we present the evaluation of typical
complicated master integrals from this class.

%\section{Results}
%A master double box with a numerator.
As has been demonstrated in \cite{GTGR0ATT},
it can be useful to use a double box with a numerator as a master integral
in order to avoid coefficients of order $1/\epsilon$ in the course
of the tensor reduction when calculating a scattering amplitude.
Therefore we present here results for the planar double box
$B_{PL,1}(1,\ldots,1,-1;s,t,m^2;\ep)$.

Our calculation of this Feynman integral is quite similar to the previous
case \cite{S3}. It is based on (\ref{6MB}). After a preliminary analysis of the
integrand one observes that it is reasonable to start the
procedure of resolution of the singularities in $\ep$, by shifting
contours and taking residues, from taking care of the gamma
function $\Gm(-1-z_2-z_3-z_4)$. Here one meets the same problem as in the massless
case (see the third reference in \cite{GTGR0ATT}) connected with
spurious singularities in MB integrals. It can also be cured in
the same way, by introducing an auxiliary analytic regularization.
To do this, we have chosen $a_8=-1+x$. Then the singularities in
the MB integrals are first resolved with respect to $x$ and then
with respect to $\ep$ when $x$ and $\ep$ tend to zero.
With this complication, the key gamma function becomes  $\Gm(x-1-z_2-z_3-z_4)$.
As a result of the procedure of resolution of the singularities in $x$ and
$\ep$, the spurious singularities in $x$ drop out, and we are left
only with six MB integrals which can be evaluated by expanding the
integrands in $\ep$ and evaluating the corresponding finite MB
integrals.

We have obtained the following result:
\bea
B_{PL,1}(1,\ldots,1,-1;s,t,m^2;\ep)
&& \nn \\ &&  \hspace*{-60mm}
= \frac{\left(i\pi^{d/2}
\e^{-\gm_{\rm E}\ep} \right)^2 x^2}{s^2 (m^2)^{2\ep}}
\left[ \frac{c^1_2 (x)}{\ep^2} + \frac{c^1_1 (x)}{\ep}
  + c^1_{01} (x)+ c^1_{02} (x)+4 \ln\frac{1+x}{1-x} c^1_{03} (x,y)
+ O(\ep) \right ]
\;,
\nn
\eea
with
\bea
c^1_2 (x) &=&  \frac{3}{2} \ln^2\frac{1-x}{1+x} \;,
\label{ResultEp2}
\\  % \hspace*{-10mm}
c^1_1 (x) &=&
\Li{3}{ \frac{(1 - x)^2}{(1 + x)^2}}
-\left( 2 \ln\frac{x}{1+x}+2\ln 2 -\ln \frac{-s}{m^2}\right)
\Li{2}{ \frac{(1 - x)^2}{(1 + x)^2}}
\nn \\ &&  \hspace*{-10mm}
+2 \ln\frac{1-x}{1+x}\, \Li{2}{\frac{1 - x}{2}}
+\frac{1}{6}\ln^3(1-x)+\frac{3}{2}\ln^2(1-x)\ln(1+x)
\nn \\ &&  \hspace*{-10mm}
-\frac{5}{2}\ln(1-x)\ln^2(1+x)+\frac{5}{6}\ln^3(1+x)
-2\,\ln 2\,\ln(1-x)\ln\frac{1-x}{1+x} \,
\nn \\ &&  \hspace*{-10mm}
-\frac{\pi^2}{6} \ln\frac{-s}{m^2}
-2\ln^2\frac{1-x}{1+x}\, \ln\frac{-t}{m^2}
-\frac{\pi^2}{6}\left(3\ln(1-x)-2\ln x\,-\ln(1+x)\right)
\nn \\ &&  \hspace*{-10mm}
+\ln^2 2\,\ln\frac{1-x}{1+x}
+\frac{\pi^2}{3} \ln 2\, -\zeta(3)\;,
\\
c^1_{01} (x) &=& 8  H\left(-1, 0, 0, 1;\frac{-1+x}{\quad 1+x}\right)\,,
%\label{ResultEp1}
\eea
where $x=1/\sqrt{1-4m^2/s}$ and $H(-1, 0, 0, 1;z)$ is a HPL
\cite{HPL}.

The two other contributions to the finite part are more
cumbersome:
\bea
c^1_{02} (x) &=&
    3 \Li{4}{(1 - x)^2/(1 + x)^2}-8 \Li{4}{(1 - x)/2} - 8 \Li{4}{-2 x/(1 - x)}
\nn \\ &&  \hspace*{-20mm}
- 12 \Li{4}{-(1 - x)/(1 + x)} -
    4 \Li{4}{(1 - x)/(1 + x)} - 8 \Li{4}{2 x/(1 + x)}
\nn \\ &&  \hspace*{-20mm}
- 8 \Li{4}{(1 + x)/2}
    - 2 S_{2,2}\left( (1 -x)^2/(1 + x)^2 \right)
    - 2 (4 l_2 - m_x - 3 p_x) \Li{3}{(1 - x)/2}
\nn \\ &&  \hspace*{-20mm}
+ 4 (m_x - p_x) \Li{3}{1 - x}
    - 2 (2 l_2 + l_x - m_x - p_x) \Li{3}{-4 x/(1 - x)^2}
\nn \\ &&  \hspace*{-20mm}
+ 4 (m_x - p_x) \Li{3}{-x/(1 - x)} +
    2 (2 l_2 + 2 l_x - 3 m_x + p_x) \Li{3}{-2 x/(1 - x)}
\nn \\ &&  \hspace*{-20mm}
- 2 (3 l_2 + l_x - m_x - 2 p_x) \Li{3}{(1 - x)^2/(1 + x)^2}
\nn \\ &&  \hspace*{-20mm}
- 2 (2 l_2 + l_x - m_x - p_x) \Li{3}{4 x/(1 + x)^2}
- 4 (m_x - p_x) \Li{3}{1/(1 + x)}
\nn \\ &&  \hspace*{-20mm}
-
    2 (6 l_2 + 2 l_x + m_x - 7 p_x) \Li{3}{(1 - x)/(1 + x)}
\nn \\ &&  \hspace*{-20mm}
- 4 (5 l_2 + 3 l_x - 2 m_x - 3 p_x) \Li{3}{-(1 - x)/(1 + x)} -
    4 (m_x - p_x) \Li{3}{x/(1 + x)}
\nn \\ &&  \hspace*{-20mm}
+ 2 (2 l_2 + 2 l_x + m_x - 3 p_x) \Li{3}{2 x/(1 + x)} -
    2 (4 l_2 - 3 m_x - p_x) \Li{3}{(1 + x)/2}
\nn \\ &&  \hspace*{-20mm}
+ 4 (l_2 + l_x - m_x) \Li{3}{-(1 + x)/(1 - x)}
+ 4 \Li{2}{(1 - x)/2}^2  + 2 \Li{2}{-x}^2
\nn \\ &&  \hspace*{-20mm}
 + (m_x - p_x) (-6 l_2 - 12 l_x + 3 m_x + 3 p_x +
          2 \ln(-s/m^2)) (\Li{2}{-x} - \Li{2}{x})
\nn \\ &&  \hspace*{-20mm}
- 4 \Li{2}{-x} \Li{2}{x}
 + (2 l_2 + l_x - 2 m_x) (m_x - p_x) \Li{2}{-4 x/(1 - x)^2}
\nn \\ &&  \hspace*{-20mm}
 + (m_x - p_x) (2 l_2 + 4 l_x - m_x - \ln(-s/m^2))
 \Li{2}{-2 x/(1 - x)}
+ (2 l_2^2 + 2 l_2 l_x
\nn \\ &&  \hspace*{-20mm}
+ l_x^2/2 - 4 l_2 m_x
- 3 l_x m_x - m_x^2 + l_x p_x + 6 m_x p_x - 3 p_x^2)
          \Li{2}{(1 - x)^2/(1 + x)^2}
\nn \\ &&  \hspace*{-20mm}
- 2 (2 l_2 + l_x - 2 p_x) (m_x - p_x)
      \Li{2}{4 x/(1 + x)^2}
- (4 l_2^2 + 4 l_2 l_x + l_x^2
- 8 l_2 m_x
\nn \\ &&  \hspace*{-20mm}
- 4 l_x m_x - m_x^2 - \pi^2/3  + 10 m_x p_x - 5 p_x^2)
\Li{2}{(1 - x)/(1 + x)}
- (4 l_2^2 + 4 l_2 l_x + l_x^2
\nn \\ &&  \hspace*{-20mm}
- 4 l_2 m_x - 2 m_x^2 - \pi^2  - 4 l_2 p_x -
          4 l_x p_x + 8 m_x p_x - 2 p_x^2
          + 4 \Li{2}{(1 - x)/(1 + x)}
\nn \\ &&  \hspace*{-20mm}
- 2 \Li{2}{-(1 - x)/(1 + x)})
      \Li{2}{-(1 - x)/(1 + x)}
\nn \\ &&  \hspace*{-20mm}
      - (m_x - p_x) (2 l_2 + 4 l_x - m_x - 2 p_x -
          \ln(-s/m^2)) \Li{2}{2 x/(1 + x)}
\nn \\ &&  \hspace*{-20mm}
+ 6 \Li{2}{-x} \Li{2}{(1 + x)/2} +
2 (-3 m_x - 4 l_x m_x + 2 m_x^2 + 3 p_x + 4 l_x p_x - 4 m_x p_x
+ 2 p_x^2
\nn \\ &&  \hspace*{-20mm}
 + 2 m_x \ln(-s/m^2) - 2 p_x \ln(-s/m^2)
- 3 \Li{2}{x} + 2 \Li{2}{(1 + x)/2})\Li{2}{(1 + x)/2}
\nn \\ &&  \hspace*{-20mm}
- 2 (3 m_x - 8 l_2 m_x
- 10 l_x m_x + 2 m_x^2 - 3 p_x + 8 l_2 p_x +
10 l_x p_x + 4 m_x p_x - 6 p_x^2
\nn \\ &&  \hspace*{-20mm}
+ 2 m_x \ln(-s/m^2) - 2 p_x \ln(-s/m^2) +
3 \Li{2}{-x} - 3 \Li{2}{x}
\nn \\ &&  \hspace*{-20mm}
+ 4 \Li{2}{(1 + x)/2}) \Li{2}{(1 - x)/2}+ 2 \Li{2}{x}^2
\nn \\ &&  \hspace*{-20mm}
 + 2 l_2^4 - 6 l_2^2 m_x + 4 l_2^3 m_x + 6 l_2^2 l_x m_x +
    6 l_2 m_x^2 - 41 l_2^2 m_x^2
\nn \\ &&  \hspace*{-20mm}
    - 32 l_2 l_x m_x^2 +
    2 l_x^2 m_x^2 + (65 l_2 m_x^3)/3 + (23 l_x m_x^3)/3
    - (49 m_x^4)/8 - (2 l_2^2 \pi^2 )/3
\nn \\ &&  \hspace*{-20mm}
+ m_x \pi^2  + (5 l_2 m_x \pi^2 )/3 - (l_x m_x \pi^2 )/2
    - (7 m_x^2 \pi^2 )/12 + (37 \pi^4 )/360
\nn \\ &&  \hspace*{-20mm}
+ 6 l_2^2 p_x - 12 l_2^3 p_x - 6 l_2^2 l_x p_x
+ 70 l_2^2 m_x p_x + 52 l_2 l_x m_x p_x
\nn \\ &&  \hspace*{-20mm}
- 4 l_x^2 m_x p_x - 6 m_x^2 p_x
+ 17 l_2 m_x^2 p_x + 13 l_x m_x^2 p_x + (4 m_x^3 p_x)/3
\nn \\ &&  \hspace*{-20mm}
- \pi^2  p_x - (l_2 \pi^2  p_x)/3 + (l_x \pi^2  p_x)/2
- (11 m_x \pi^2  p_x)/6 - 6 l_2 p_x^2
\nn \\ &&  \hspace*{-20mm}
- 17 l_2^2 p_x^2
- 20 l_2 l_x p_x^2 + 2 l_x^2 p_x^2 + 6 m_x p_x^2 - 87 l_2 m_x p_x^2
- 43 l_x m_x p_x^2
\nn \\ &&  \hspace*{-20mm}
- (41 m_x^2 p_x^2)/4 + (7 \pi^2  p_x^2)/4
+ (121 l_2 p_x^3)/3 + (67 l_x p_x^3)/3 + (113 m_x p_x^3)/3
\nn \\ &&  \hspace*{-20mm}
- (165 p_x^4)/8 - 2 (2 l_2 + 4 l_x - m_x - p_x) (m_x - p_x)^2 \ln(-s/m^2)
\nn \\ &&  \hspace*{-20mm}
+ 2 (m_x - p_x)^2 \ln^2(-s/m^2) + 2 (m_x - p_x)^2 \ln^2(-t/m^2)
\nn \\ &&  \hspace*{-20mm}
+ 12 l_2 \zeta_3 + 6 l_x \zeta_3  - 4 m_x \zeta_3
- 8 p_x \zeta_3\,,
\eea
%%%%%%%%%%%%%%%%%%%%%%%%%%%%%%%%%%%%%%%%%%%%%%%%%%%%%%%%%%%%
\bea
c^1_{03} (x,y) &=&
2(\Li{3}{(1 - x)/2} - \Li{3}{(1 + x)/2}
+ \Li{3}{(1 - x) y/(1 - x y)}
\nn \\ &&  \hspace*{-24mm}
- \Li{3}{-(1 + x) y/(1 - x y)}
+ \Li{3}{-(1 - x) y/(1 + x y)} - \Li{3}{(1 + x) y/(1 + x y)})
\nn \\ &&  \hspace*{-24mm}
+ \Li{3}{(1 + x) (1 - y)/(2(1 - x y))}
- \Li{3}{(1 - x) (1 + y)/(2(1 - x y))}
\nn \\ &&  \hspace*{-24mm}
- \Li{3}{(1 - x) (1 - y)/(2(1 + x y))}
+ \Li{3}{(1 + x) (1 + y)/(2(1 + x y))}
\nn \\ &&  \hspace*{-24mm}
+ 2 (l_2 - m_x) \Li{2}{(1 - x)/2} + 2 (p_x-l_2) \Li{2}{(1 + x)/2}
- (m_{xy} - p_{xy}) \Li{2}{(1 - y)/2}
\nn \\ &&  \hspace*{-24mm}
+ 2 (m_{xy} - p_{xy}) (\Li{2}{-y} - \Li{2}{y})
+ (m_{xy} - p_y) \Li{2}{((1 + x) y)/(1 + y)}
\nn \\ &&  \hspace*{-24mm}
+ (m_{xy} - p_{xy}) \Li{2}{(1 + y)/2}
+ (m_{xy} - p_y) \Li{2}{(1 - x y)/(1 + y)}
\nn \\ &&  \hspace*{-24mm}
- (p_{xy} - p_y) \Li{2}{(y - x y)/(1 + y)}
- 2 (l_y + m_x - m_{xy}) \Li{2}{(1 - x) y/(1 - x y)}
\nn \\ &&  \hspace*{-24mm}
+ 2 (l_y - m_{xy} + p_x) \Li{2}{-(1 + x) y/(1 - x y)}
\nn \\ &&  \hspace*{-24mm}
-  2 (l_y + m_x - p_{xy}) \Li{2}{-(1 - x) y/(1 + x y)}
+  2 (l_y + p_x - p_{xy}) \Li{2}{(1 + x) y/(1 + x y)}
\nn \\ &&  \hspace*{-24mm}
- (p_{xy} - p_y) \Li{2}{(1 + x y)/(1 + y)}
\nn \\ &&  \hspace*{-24mm}
+ (l_2 + m_{xy} - m_y - p_x) \Li{2}{(1 + x) (1 - y)/(2(1 -  x y))}
\nn \\ &&  \hspace*{-24mm}
- (l_2 - m_x + m_{xy} - p_y) \Li{2}{(1 - x) (1 + y)/(2(1 -  x y))}
\nn \\ &&  \hspace*{-24mm}
- (l_2 - m_x - m_y + p_{xy}) \Li{2}{(1 - x) (1 - y)/(2(1 +  x y))}
\nn \\ &&  \hspace*{-24mm}
+ (l_2 - p_x + p_{xy} - p_y) \Li{2}{(1 + x) (1 + y)/(2(1 +  x y))}
\nn \\ &&  \hspace*{-24mm}
- (m_{xy} - m_y + p_{xy} - p_y)
(2\Li{2}{x} - 2\Li{2}{-x} + \Li{2}{-2 x/(1 - x)}
\nn \\ &&  \hspace*{-24mm}
- \Li{2}{2 x/(1 + x)})
+ (2 m_x m_{xy} - 2 m_x m_y - m_{xy} m_y
- 2 m_{xy} p_x + 2 m_y p_x
\nn \\ &&  \hspace*{-24mm}
 + 2 m_x p_{xy} + m_y p_{xy}
- 2 p_x p_{xy} - 2 m_x p_y + m_{xy} p_y + 2 p_x p_y - p_{xy} p_y) l_2
+ 2 l_y m_x m_{xy}
\nn \\ &&  \hspace*{-24mm}
+ l_y m_{xy}^2 - m_x m_{xy}^2
- 2 l_y m_x m_y + m_x m_{xy} m_y
- m_{xy}^2 m_y + (m_{xy} m_y^2)/2
\nn \\ &&  \hspace*{-24mm}
- 2 l_y m_{xy} p_x
+ 2 m_{xy}^2 p_x + 2 l_y m_y p_x + m_{xy} m_y p_x
\nn \\ &&  \hspace*{-24mm}
+ (5 \pi^2  (m_{xy} - p_{xy}))/6 + 2 l_y m_x p_{xy}
- m_x m_y p_{xy} - (m_y^2 p_{xy})/2 - 2 l_y p_x p_{xy}
\nn \\ &&  \hspace*{-24mm}
- m_y p_x p_{xy} - l_y p_{xy}^2 - 2 m_x p_{xy}^2 + m_y p_{xy}^2
+ p_x p_{xy}^2 - 2 l_y m_x p_y - 2 l_y m_{xy} p_y - m_x m_{xy} p_y
\nn \\ &&  \hspace*{-24mm}
+ 2 m_x m_y p_y + 2 l_y p_x p_y - 3 m_{xy} p_x p_y - 2 m_y p_x p_y
+ 2 l_y p_{xy} p_y + 3 m_x p_{xy} p_y + p_x p_{xy} p_y
\nn \\ &&  \hspace*{-24mm}
- m_x p_y^2
+ (3 m_{xy} p_y^2)/2 +  p_x p_y^2 - (3 p_{xy} p_y^2)/2\,,
\eea
where $ y=1/\sqrt{1-4m^2/t}$ and $S_{2,2}$ is a generalized
polylogarithm \cite{GenPolyLog}.
The following abbreviations are also used: $\zeta_3=\zeta(3)$,
$l_z =\ln z$ for $z=x,y,2$,  $p_z =\ln(1+z)$ and
$m_z =\ln(1-z)$ for $z=x,y,xy$.

%Numerical control of finite MB integrals.
%Numerical confirmation by the method of  \cite{BH}.
The results have been checked numerically in two independent ways:
The finite MB integrals, obtained after the resolution of singularities,
have been evaluated by numerical integration along the imaginary axis.
The parameter space for the  Monte Carlo integration was maximally
4-dimensional in this case.
In addition, the overall coefficients of the poles
have been calculated at Euclidean points by the completely
independent method of sector decomposition\,\cite{BH,baba}.

\subsection{Planar double boxes of the second type}

For the general planar double box of the second type (see Fig.~1b)
\bea
B_{PL,2}(a_1,\ldots,a_8;s,t,m^2;\ep) &=&
\int\int \frac{\dd^dk \, \dd^dl}{(k^2-m^2)^{a_1}[(k+p_1)^2)]^{a_2}
[(k+p_1+p_2)^2-m^2]^{a_3}}
\nn \\ && \hspace*{-50mm}
\times \frac{[(k+p_1+p_2+p_3)^2]^{-a_8}}{
[(l+p_1+p_2)^2)]^{a_4}[(l+p_1+p_2+p_3)^2-m^2]^{a_5}
(l^2)^{a_6} [(k-l)^2-m^2]^{a_7} }
\, ,
\label{2box-PL2}
\eea
a sixfold MB representation can be derived \cite{S5} similarly to the
first case:
\bea
B_{PL,2}(a_1,\ldots,a_8;s,t,m^2;\ep)
&& \nn \\ &&  \hspace*{-53mm}
=\frac{\left(i\pi^{d/2} \right)^2 (-1)^a}{
\prod_{j=2,4,5,6,7}\Gm(a_j) \Gm(4-a_{4567}-2\ep)(-s)^{a-4+2\ep}}
\nn \\ &&  \hspace*{-53mm}\times
\frac{1}{(2\pi i)^6} \int_{-i\infty}^{+i\infty}
\prod_{j=1}^6 \dd z_j
\left(\frac{m^2}{-s} \right)^{z_5+z_6}
\left(\frac{t}{s} \right)^{z_1}
\prod_{j=1}^6 \Gm(-z_j)
%\Gm(-z_1)\Gm(-z_2) \Gm(-z_3)\Gm(-z_4)\Gm(-z_5)\Gm(-z_6)
\frac{
\Gm(a_2 + z_1) \Gm(a_4 + z_2 + z_4)}
{\Gm(a_3 - z_2)\Gm(a_1 - z_3)
}
\nn \\ &&  \hspace*{-53mm}\times
\frac{ \Gm(4 - a_{445667} - 2 \ep - z_2 - z_3 - 2 z_4) \Gm(a_6 + z_3 + z_4)
}
{\Gm(4 - a_{445667} - 2 \ep -  z_2 - z_3 - 2 z_4 - 2 z_5)
\Gm(6 -a - 3 \ep - z_4 - z_5)
}
\nn \\ &&  \hspace*{-53mm}\times
\frac{
   \Gm(8 - a_{13}-2 a_{245678} - 4 \ep
  - 2 z_1 - z_2 - z_3 - 2 z_4 - 2 z_5)}
{\Gm(8 - a_{13} - 2 a_{245678} - 4 \ep
  - 2 z_1 - z_2 - z_3 - 2 z_4 - 2 z_5 - 2 z_6)}
\nn \\ &&  \hspace*{-53mm}\times
\frac{\Gm(2 - a_{456} - \ep - z_4 - z_5)
\Gm(2 - a_{467} - \ep - z_2 - z_3 - z_4 - z_5)
  }
{\Gm(a_{45678}-2 +\ep + z_2 + z_3 + z_4 + z_5) }
\nn \\ &&  \hspace*{-53mm}\times
 \Gm( a_{4567} +\ep -2  + z_2 + z_3 + z_4 + z_5)
\Gm(a_{45678}-2 +\ep + z_1 + z_2 + z_3 + z_4 + z_5)
\nn \\ &&  \hspace*{-53mm}\times
\Gm(4 - a_{1245678} - 2 \ep -z_1 - z_2 - z_4 - z_5 - z_6)
\nn \\ &&  \hspace*{-53mm}\times
\Gm(4 - a_{2345678} - 2 \ep - z_1 - z_3 - z_4 - z_5 - z_6)
\nn \\ &&  \hspace*{-53mm}\times
\Gm( a -4+ 2 \ep + z_1 + z_4 + z_5 + z_6)
    \, ,
\label{6MB-PL2}
\eea
%(where $a_{jjk\ldots}=2a_j+a_k+\ldots$)\,.

We have used this representation to calculate
the master planar double box of the second type
$B_{PL,2}(1,\ldots,1,0;s,t,m^2;\ep)$.
The resolution of singularities in $\ep$ was performed similar to
the previous cases. The number of resulting MB integrals where an
expansion in $\ep$ can be performed in the integrand is again
equal to six. We have arrived at the following result:
\bea
B_{PL,2}(1,\ldots,1,0;s,t,m^2;\ep)
&=&\left(i\pi^{d/2}
\e^{-\gm_{\rm E}\ep} \right)^2
\left[
-\frac{x y}{s^2 (-t)^{1+2\ep}}
\right.  \nn \\ &&  \hspace*{-60mm}
\times \left(
\frac{c^2_2 (x,y)}{\ep^2} + \frac{c^2_1 (x,y)}{\ep}
  + c^2_{01} (x,y)+ c^2_{02} (x,y)
\right)
%\nn \\ &&  \hspace*{-0mm}
\left.
  + c^2_{03} (s,t,m^2)
+ O(\ep) \right ]
\;,
\eea
where
\bea
c^2_2(x) &=& \ln\frac{1-x}{1+x} \ln\frac{1-y}{1+y}
\;,
%\label{ResultEp2}
\\  % \hspace*{-10mm}
c^2_1 (x) &=&
 -2  \ln\frac{1-y}{1+y}\left[
\Li{2}{\frac{1 - x}{2}} - \Li{2}{\frac{1 + x}{2}}
+\Li{2}{x}-\Li{2}{-x}\right]
\nn \\ &&  %\hspace*{-10mm}
+2 \ln\frac{1-x}{1+x}\left[
\Li{2}{\frac{1 - y}{2}} - \Li{2}{\frac{1 + y}{2}}
+\Li{2}{y}-\Li{2}{-y}\right]
\nn \\ &&  %\hspace*{-10mm}
+ \ln\frac{1-x}{1+x}\ln\frac{1-y}{1+y}\left[4\ln\frac{y}{x}
+\ln\frac{1-x}{1-y} +\ln\frac{1+x}{1+y}\right]
\;.
\eea
Furthermore,
\bea
c^2_{01} (x,y) &=&
4 ((-m_y + p_y) (3 \Li{3}{(1 - x)/2} + \Li{3}{1 - x}
     + \Li{3}{-x/(1 - x)}
\nn \\ &&  \hspace*{-24mm}
     - \Li{3}{-2 x/(1 - x)}
%\nn \\ &&  \hspace*{-24mm}
     - \Li{3}{1/(1 + x)}
     - \Li{3}{x/(1 + x)}
     + \Li{3}{2 x/(1 + x)}
\nn \\ &&  \hspace*{-24mm}
     - 3 \Li{3}{(1 + x)/2})
+ (m_x -p_x) (\Li{3}{(1 - y)/2} - \Li{3}{1 - y}
- 2 \Li{3}{-y} + 2 \Li{3}{y}
\nn \\ &&  \hspace*{-24mm}
- \Li{3}{-y/(1 - y)}
+ \Li{3}{-2 y/(1 - y)} + \Li{3}{1/(1 + y)}
+ \Li{3}{y/(1 + y)}
\nn \\ &&  \hspace*{-24mm}
- \Li{3}{2 y/(1 + y)}
- \Li{3}{(1 + y)/2}))
+ 4 \ln(-s/m^2) ((m_y - p_y) (\Li{2}{(1 - x)/2}
\nn \\ &&  \hspace*{-24mm}
- \Li{2}{(1 + x)/2}) + (m_x - p_x) (\Li{2}{-y}
- \Li{2}{y}))
\nn \\ &&  \hspace*{-24mm}
+ 2 ((p_y-m_y ) ((2 l_2 + 4 l_y - m_y - p_y) \Li{2}{x}
+ 2 (l_2 + l_x - m_x) \Li{2}{-2 x/(1 - x)}
\nn \\ &&  \hspace*{-24mm}
- 2 (l_2 + l_x - p_x) \Li{2}{(2 x)/(1 + x)}
\nn \\ &&  \hspace*{-24mm}
- (2 + 2 l_2 + 4 l_x + 4 l_y - m_y - 4 p_x - p_y) \Li{2}{(1 + x)/2})
\nn \\ &&  \hspace*{-24mm}
+ (m_x - p_x) (2 - 2 l_2 - 4 l_x + m_x
- 2 m_y + p_x + 2 p_y) \Li{2}{(1 - y)/2}
\nn \\ &&  \hspace*{-24mm}
- 2 \Li{2}{x} \Li{2}{(1 - y)/2}
+ 2 \Li{2}{(1 + x)/2} \Li{2}{(1 - y)/2}
\nn \\ &&  \hspace*{-24mm}
+ ((m_x - p_x) (2 l_2 - 4 l_x + m_x + p_x) +
                2 \Li{2}{x} + 2 \Li{2}{(1 + x)/2})
            \Li{2}{-y}
\nn \\ &&  \hspace*{-24mm}
- ((m_x - p_x) (2 l_2 - 4 l_x + m_x + p_x) +
                2 \Li{2}{x} + 2 \Li{2}{(1 + x)/2}) \Li{2}{y}
\nn \\ &&  \hspace*{-24mm}
- 2 (m_x - p_x) ((l_2 + l_y - m_y) \Li{2}{-2 y/(1 - y)}
- (l_2 + l_y - p_y) \Li{2}{(2 y)/(1 + y)})
\nn \\ &&  \hspace*{-24mm}
+ \Li{2}{-x} (2 l_2 m_y + 4 l_y m_y - m_y^2 - 2 l_2 p_y - 4 l_y p_y
+ p_y^2 + 2 \Li{2}{(1 - y)/2} - 2 \Li{2}{-y}
\nn \\ &&  \hspace*{-24mm}
+ 2 \Li{2}{y}
- 2 \Li{2}{(1 + y)/2})
- ((m_x - p_x) (-2 + 2 l_2 - 4 l_x + 4 l_y
+ m_x + p_x - 4 p_y)
\nn \\ &&  \hspace*{-24mm}
- 2 \Li{2}{x}
+ 2 \Li{2}{(1 + x)/2}) \Li{2}{(1 + y)/2}
\nn \\ &&  \hspace*{-24mm}
+  \Li{2}{(1 - x)/2} ((m_y - p_y)
(2 - 6 l_2 - 8 l_x - 4 l_y + 6 m_x + m_y + 2 p_x + p_y)
\nn \\ &&  \hspace*{-24mm}
- 2 \Li{2}{(1 - y)/2} - 2 \Li{2}{-y}
+ 2 \Li{2}{y} + 2 \Li{2}{(1 + y)/2}))
\nn \\ &&  \hspace*{-24mm}
- (m_x - p_x) (m_y - p_y) \ln^2(-s/m^2)
+ (m_x m_y (16 l_2 ( l_x + l_y-1) + \pi^2 ))/2
\nn \\ &&  \hspace*{-24mm}
+ (2 (4 l_x - m_x - p_x) (m_x - p_x) (m_y - p_y) \ln(-s/m^2))
\nn \\ &&  \hspace*{-24mm}
+ (8 m_y^3 (m_x - p_x))/3 - 6 l_2 m_x^2 (m_y - p_y)
+ (4 m_x^3 (m_y - p_y))/3 - 8 l_x^2 (m_x - p_x) (m_y - p_y)
\nn \\ &&  \hspace*{-24mm}
+  8 l_y^2 (m_x - p_x) (m_y - p_y) + 2 (l_2 + m_x) p_x^2 (m_y - p_y)
- 2 p_x^3 (m_y - p_y)
\nn \\ &&  \hspace*{-24mm}
- 4 l_2^3 (m_x + m_y - p_x - p_y)
+ (2 (l_2-1) \pi^2  (m_x + m_y - p_x - p_y))/3
\nn \\ &&  \hspace*{-24mm}
+ (6 m_x m_y - 2 l_2 (m_x - p_x)
- p_x (6 m_y + p_x)) p_y^2
- (10 (m_x - p_x) p_y^3)/3
\nn \\ &&  \hspace*{-24mm}
- m_y^2 (m_x - p_x) (2 l_2 + 8 l_y + m_x + p_x + 2 p_y)
\nn \\ &&  \hspace*{-24mm}
+ (\pi^2  (m_y p_x + (m_x - 5 p_x) p_y))/6
+ m_x^2 (2 m_y p_x - 2 p_x p_y + p_y^2)
\nn \\ &&  \hspace*{-24mm}
+ (2 l_y (m_x - p_x) (\pi^2  + 6 m_y p_x + 6 m_x (m_y - p_y)
- 6 m_y p_y - 6 p_x p_y + 12 p_y^2))/3
\nn \\ &&  \hspace*{-24mm}
- 4 l_2^2 (p_x - l_y p_x + m_y (l_x + p_x-1 )
+ p_y - l_x p_y + 2 p_x p_y + m_x (-1 + l_y - 4 m_y + p_y))
\nn \\ &&  \hspace*{-24mm}
+ (2 l_x (m_y - p_y) (\pi^2  - 24 l_y (m_x - p_x) - 6 m_y p_x
- 6 p_x p_y + 6 m_x (m_y - p_x + p_y)))/3
\nn \\ &&  \hspace*{-24mm}
- 8 l_2 (l_y m_y p_x - (1 + m_y) p_x p_y
+ m_x ((l_x - p_x) p_y + m_y (p_x + p_y)))
\nn \\ &&  \hspace*{-24mm}
+ 4 (-(m_y p_x p_y) + m_x (-(p_x p_y) + m_y (p_x + p_y)))\,.
\eea
The function $c^2_{02} (x,y)$ is equal to $(p_x - m_x) \bar{b}_{02}(y, x)$
where the function  $\bar{b}_{02}(x, y)$ is given
by the braces in Eq.~(9) of \cite{S3} which defines
the contribution $b_{02}(x, y)$ to the finite part of the
master double box of the first type.

Finally,
\bea
c^2_{03} (s,t,m^2)&=&
-\frac{4}{s\sqrt{-t}}
\int_0^1\int_0^1\dd x_1 \dd x_2
\frac{\sqrt{x_1}}{\sqrt{1 - x_1}}
\nn \\ &&  \hspace*{-10mm}
\times
\frac{
 \mbox{Arsh}
\left[(\sqrt{-t} \sqrt{x_1} \sqrt{1 - x_2})/
    (2 m \sqrt{x_1 + x_2 - x_1 x_2})\right]}
{ (4 m^2 -s x_1) x_2 \sqrt{
(4 m^2 -t) x_1 (1 - x_2) + 4 m^2 x_2}}
\nn \\ &&  \hspace*{-10mm}
%\times (\ln(-s/m^2) + 2\ln x_1 -
%   \ln\left(-s x_1^2 + (1 - x_1) (4 -s x_1) x_2\right)
\times (\ln(-s/m^2) + 2\ln x_1 -
\ln\left[
4(1-x_1)x_2-s x_1(x_1+x_2-x_1 x_2)/m^2
\right]
%
%\nn \\ &&  %\hspace*{-10mm}
%\int_0^1\int_0^1\dd x_1 \dd x_2\frac{
% \ln^2\left(x_1 + x_2 - x_1 x_2\right)}{
%\sqrt{1 - x_1}(4 - s x_1)  \sqrt{1 - x_2} (4 -t x_2)}
%\,,
\nn \\ &&  \hspace*{-10mm}
+ \frac{1}{s}
\int_0^1\int_0^1\dd x_1 \dd x_2\frac{
  \ln(x_1 + x_2 - x_1 x_2)}
{\sqrt{1 - x_1} (4m^2 -s x_1) \sqrt{1 - x_2} (4 m^2 -t x_2)}
\nn \\ &&  \hspace*{-10mm}
\times
 \left(4 \ln 2 + 2\ln(-4 m^2/s +x_1) + 2\ln(1 - x_1)
\right.
\nn \\ &&  \hspace*{-10mm}
\left.
-2 \ln x_1 + 2\ln(1 - x_2)
 - \ln(x_1 + x_2 - x_1 x_2) \right)\,.
\eea

We have controlled this result similarly to the previous case, by
numerical evaluation of our finite MB integrals and numerical
evaluation by the method of \cite{BH}.

It is not clear whether the two-parametric integrals present in the
finite part can be expressed in terms of
HPL or 2dHPL depending on special combinations
of $s,t$ and $m^2$.
If the answer to this question is negative one might think about
the introduction of  a new class of functions, e.g., a sort of
generalized two-dimensional HPL. In this context, let us point
out that generalized HPL of various types were introduced in
\cite{appl-vert2}.
These new special functions were defined similarly
to HPL, with other
basic functions, in particular $1/\sqrt{t(t+4)}$.

However, for example, the GHPL
\[
H(-r, -1; x)= \int_0^x \frac{\dd t}{\sqrt{t(t+4)}}
\]
defined in \cite{appl-vert2} equals
\[
2 \mbox{Li}_{2}\left(z,\frac{\pi}{3} \right)
+\frac{1}{2}\ln ^2 z -\frac{\pi^2}{18}\,,
\]
where
\[
z=\frac{\sqrt{4 + x}-\sqrt{x}} {\sqrt{4 + x}-\sqrt{x}}
\]
and
$\mbox{Li}_{2}\left(r, \theta \right)=\mbox{Re}\left[
\mbox{Li}_{2}\left(r e^{i \theta} \right)\right]$
is the dilogarithm of a complex argument.
Still it is not yet clear whether any GHPL can be expressed in
terms of polylogarithms and HPL.

From the mathematical point of view it is natural to try to
express any new results in terms of known functions, which may depend
on the initial variables through some special combinations.
Pragmatically, it is sufficient to present new results in terms of
some functions which are, probably, new and which can be
evaluated with high precision at physical values of the given
variables.
Anyway, the two-parametrical integrals presented above can be
evaluated numerically in a straightforward way at any non-singular point
of the variables $s,t,m^2$.

\subsection{Non-planar double boxes}

For the general non-planar double box (see Fig.~1c),
one can derive the following eightfold MB representation
\cite{S5}:
\bea
B_{NP}(a_1,\ldots,a_8;s,t,u,m^2;\ep)
 &=&
\frac{\left(i\pi^{d/2} \right)^2 (-1)^a}{
\prod_{j=2,4,5,6,7}\Gm(a_j) \Gm(4-a_{4567}-2\ep)(-s)^{a-4+2\ep}}
\nn \\ &&  \hspace*{-68mm}\times
\frac{1}{(2\pi i)^8} \int_{-i\infty}^{+i\infty}
\prod_{j=1}^8 \dd z_j
\left(\frac{m^2}{-s} \right)^{z_5+z_6}
\left(\frac{t}{s} \right)^{z_7}\left(\frac{u}{s} \right)^{z_8}
\prod_{j=1}^7 \Gm(-z_j)
\nn \\ &&  \hspace*{-68mm}
\frac{\Gm(a_5 + z_2 + z_4)
 \Gm(a_7 + z_3 + z_4)
\Gm(4 - a_{455677} - 2 \ep - z_2 - z_3 -  2 z_4)
 }
{\Gm(a_1 - z_2) \Gm(a_3 - z_3) \Gm(a_8 - z_4)
}
\nn \\ &&  \hspace*{-68mm}
\times
\frac{
  \Gm(2 - a_{567} - \ep - z_2 - z_4 - z_5)
  \Gm(2 - a_{457} - \ep - z_3 - z_4 - z_5)
}
{\Gm(4 - a_{455677} - 2 \ep - z_2 - z_3 - 2 z_4 - 2 z_5)
}
\nn \\ &&  \hspace*{-68mm}
\times
\frac{ \Gm(a_8 + z_1 - z_4 + z_7)\Gm(8 - a_{13} - 2 a_{245678} - 4 \ep
- z_2 - z_3 - 2 z_5 - 2 z_7 - 2 z_8)
  }
{\Gm(6 - a - 3 \ep - z_5) }
\nn \\ &&  \hspace*{-68mm}
\times
\frac{ \Gm(-a_8 - z_1 + z_4 - z_7 - z_8)\Gm(4 - a_{2345678} - 2 \ep - z_2 - z_5 -z_6 - z_7 - z_8)
}
{\Gm(8 - a_{13} - 2 a_{245678} -4 \ep
- z_2 - z_3 - 2 z_5 - 2 z_6 - 2 z_7 - 2 z_8) }
\nn \\ &&  \hspace*{-68mm}
\times
\frac{\Gm(4 - a_{1245678} - 2 \ep - z_3 - z_5 - z_6 - z_7 - z_8)
\Gm(a_{28} + z_1 - z_4 + z_7 + z_8)}
{\Gm(a_{245678} -2+ \ep + z_1 + z_2 + z_3 + z_5 + z_7 + 2 z_8)}
\nn \\ &&  \hspace*{-68mm}
\times
\Gm(a_{4567} + \ep-2 + z_2 + z_3 + z_4 + z_5 + z_8)
\Gm(a-4 + 2 \ep + z_5 +z_6 + z_7 + z_8)
\nn \\ &&  \hspace*{-68mm}
\times
\Gm(a_{245678} -2+ \ep +z_1 + z_2 + z_3 + z_5 + 2 z_7 + 2 z_8)
    \, .
\label{8MB}
\eea
Although the number of integrations is rather high one can proceed
also in this case. However, it turns out that the massive non-planar
case is rather complicated.
We have performed the resolution of singularities
for the non-planar master planar double box.
Here is our result for its double-pole part in $\ep$:
\[
B_{NP}(1,\ldots,1,0;s,t,m^2;\ep)
%&&\nn \\ &&  \hspace*{-60mm}
= \frac{\left(i\pi^{d/2}\right)^2 x z}{s t u}
\ln\frac{1-x}{1+x}\,\ln\frac{1-z}{1+z}\;
 \frac{1}{\ep^2} + O\left(\frac{1}{\ep}\right)
\;,
\]
where $z=1/\sqrt{1-4m^2/u}$.
A one-parametric integral which can hardly be expressed in terms
of known special functions is present already in the $1/\ep$ part.

Unfortunately, we are unable to check our preliminary result (for
the coefficient of the simple pole and for the finite part)
numerically by the method of \cite{BH} for the following reason.
As in the case of the massless non-planar double boxes (see the
second reference in \cite{MB}), it is natural to treat the
Mandelstam variables as not restricted by the physical condition
$s+t+u=4m^2$, because this condition does not simplify the
calculation. Then the natural procedure is to perform the
calculation for some extension of the given Feynman integral as a
function of the variables $s$, $t$, $u$ (and, now, $m^2$) to the
Euclidean domain, with $s,t,u<0$, and perform the analytical
continuation to the physical domain in the result. When one starts
from the alpha/Feynman parameters of the whole diagram one
naturally arrives at an extension of the physical condition just
by considering $u$ as an independent variable. Since, in the
massless case, a fourfold MB representation was derived starting
from the global Feynman parameter representation, the same extension
as in \cite{BH} was implied there. In the massive case however, 
if we start from Feynman parameters and try to
separate various terms entering functions present in the integrals
over Feynman parameters, we do not see any possibility to arrive at a
MB representation with the number of integrations less than ten.
The eightfold MB representation (\ref{8MB}) was derived in another
way, by introducing some Feynman parameters upon integrating over
the first loop momentum, then integrating over the second loop
momentum and completing the procedure of introducing MB
integrations. After this procedure, the variable $u$ is considered
as an independent variable. It turns out that this extension to
non-physical Euclidean points defined by (\ref{8MB}) differs from
the version implied within \cite{BH}, based on the global
Feynman parametric representation for the initial diagram. This
can be seen by checking (\ref{8MB}) in limiting cases where
some of the indices $a_i$ are zero and where one has explicit analytical
results:  
In some of these limits, it is necessary to use 
the physical condition $s+t+u=4m^2$ to prove agreement with the known
result, while the agreement is achieved without this condition
when one starts from the global Feynman parametric representation.

Thus, to perform numerical control of our results,
a generalization of the method of sector decompositions \cite{BH}
to points with kinematical invariants of different signs is
necessary and will be developed in the near future. Then one will
be able to make reliable results for massive non-planar double
boxes. \vspace{0.5cm}

{\em Acknowledgments.}
The work of V.A.S. was
supported by  Volkswagen Foundation
(Contract No. I/77788) and DFG Mercator visiting professorship
No. Ha 202/110-1.
The work of G.H. was supported 
by the Bundesministerium f\"ur Bildung und Forschung
through Grant No.\ 05~HT4GUA/4.

\end{document}